\documentclass[12pt]{article}
\usepackage{graphicx}

\begin{document}

\title{Central configurations for the planar Newtonian Four-Body problem}

\author{E. Pi\~na and P. Lonngi\\
Department of Physics \\ Universidad Aut\'onoma Metropolitana - Iztapalapa, \\
P. O. Box 55 534 \\ Mexico, D. F., 09340 Mexico \\
e-mail: pge@xanum.uam.mx and plov@xanum.uam.mx}

\date{ }
\maketitle

\abstract{The plane case of central configurations with four different masses is analyzed theoretically and is computed numerically. We follow Dziobek's approach to four body central configurations with a direct implicit method of our own in which the fundamental quantities are the quotient of the directed area divided by the corresponding mass and a new simple numerical algorithm is developed to construct general four body central configurations. This tool is applied to obtain new properties of the symmetric and non-symmetric central configurations. The explicit continuous connection between three body and four body central configurations where one of the four masses approaches zero is clarified. Some cases of coorbital 1+3 problems are also considered.
}
\


\

\newpage

\section{Introduction}
In this paper we study the central configurations that appear as particular solutions of the differential equations of motion of the Newtonian 4-Body Problem, namely \cite{js}
\begin{equation}
m_i \frac{d^2}{d t^2} {\bf r}_i = \sum_{j \neq i}^4 \frac{G m_i m_j ({\bf r}_j -{\bf r}_i)}{r_{ij}^3} \quad \mbox{(for i = 1, 2, 3, 4,)} \label{ne}
\end{equation}
where ${\bf r}_j$, $(j = 1, 2, 3, 4)$ denote the position vectors of the particles in three dimensional space; $r_{ij}=r_{ji}$ are the distances between particles $i$ and $j$; $m_j$, $(j = 1, 2, 3, 4,)$ denote the constant masses, and G is the (measured) gravitational constant. The masses of the four bodies $m_1$, $m_2$, $m_3$ and $m_4$ are generally different.
The right hand side of this equation is the gravitational force which is derived from the potential energy which is the Newtonian function of the distances between particles $r_{ij}$,
\begin{equation}%
V = - \sum_{i<j}^4\frac{G m_i m_j}{r_{ij}}\, .
\end{equation}

For any configuration to be central, vector forces must be in the direction of the position vector so that the right hand side of equation (\ref{ne}) divided by the mass $m_i$ equals a constant by the position vector ${\bf r}_i$.

The relevance of central configurations is remarked in multiple sites in the literature. We appreciate specially the beautiful presentation of the general theme by D. Saari \cite{sa} where the fundamental references are found. Central configurations are the only known cases where the differential equations of the N-Body Problem are integrable. They are relevant for the simultaneous collision of the N bodies, and in the limit that the relative positions of N colliding particles increase without bound. Topological reasons ask for their properties to be discovered. Even the number of essentially different central configurations is a mathematical challenge for this century. An additional new reason is their usefulness to demonstrate the existence of non integrability of the N-Body differential equations \cite{al}. The original sources may be found in these two references.

The study of central configurations of the N-Body Problem is a subject that expands in many important directions. Since central configurations of the Three-Body Problem are better known, we tackle the Four-Body Problem.

The Four-Body Problem in the plane was already considered by some researchers in the Nineteenth Century. Again we cite Albouy et al. \cite{al} to find some of them. In 1900 Dziobek \cite{dz} published a seminal work on central configurations with relative distances as coordinates. After him remarkable papers were published, among which we select the MacMillan and Barker study \cite{mb}, with different coordinates and similar results. Later Schmidt \cite{ds} reviews and extends these results for five particles in three dimensions. Moeckel \cite{mo} proves the finiteness of the number of Dziobek's configurations. Leandro \cite{le} refines the finiteness problem and studies important bifurcations.

Although the ideal for physicists and mathematicians is the general case of four different masses, the difficulties in simplifying it have lead to a prudent research based on many particular cases with some symmetry. The case of four equal masses was analyzed by Albouy \cite{ay}. When three of the masses are equal, but different from the fourth mass, a research by Bernat, Llibre and P\'erez-Chavela \cite{be} settles interesting theorems on the number of solutions and bifurcation values in the case of kite symmetry. A reciprocal result to this is found by Long and Sun \cite{ls}. Properties of symmetric cases are also obtained by Albouy and coworkers \cite{al}.

Our main contribution to the Central Configurations literature is to set forth the four weighted directed areas as known parameters. Although the approach of not choosing the masses as fundamental data is against tradition, it is justified by the simplicity of the resulting algorithm that produces as output values for the six distances and the four masses in a compatible central configuration.

This approach leads us to one non-trivial algebraic equation in a single unknown, to be compared  with the many unknown variables approach used by others.

 \section{The plane problem}
The directed vectors representing twice the area of each of the four faces of a tetrahedron with vertices ${\bf r}_1$, ${\bf r}_2$, ${\bf r}_3$, ${\bf r}_4$, are:
 \begin{eqnarray}
{\bf S}_1 &=& {\bf r}_4 \wedge {\bf r}_2 + {\bf r}_2 \wedge {\bf r}_3 + {\bf r}_3 \wedge {\bf r}_4 \\
{\bf S}_2 &=& {\bf r}_1 \wedge {\bf r}_4 + {\bf r}_4 \wedge {\bf r}_3 + {\bf r}_3 \wedge {\bf r}_1\\
{\bf S}_3 &=& {\bf r}_1 \wedge {\bf r}_2 + {\bf r}_2 \wedge {\bf r}_4 + {\bf r}_4 \wedge {\bf r}_1 \\
{\bf S}_4 &=& {\bf r}_2 \wedge {\bf r}_1 + {\bf r}_3 \wedge {\bf r}_2 + {\bf r}_1 \wedge {\bf r}_3
\end{eqnarray}

The vectorial sum of these four vectors is the zero vector
\begin{equation}
{\bf S}_1 + {\bf S}_2 + {\bf S}_3 + {\bf S}_4 = {\bf 0}
\end{equation}
The case where the four particles at the vertices are in a constant plane is an important and old subject. The four vector directed areas are all parallel, the direction of which may be taken as the unit vector $\bf k$.
\begin{equation}
{\bf S}_j  = {\bf k} S_j\, ,
\end{equation}
where the $S_j$ are twice the directed areas, positive if ${\bf S}_j$ is parallel to $\bf k$, and negative if the direction of ${\bf S}_j$ is opposite to $\bf k$.

We choose the third component of the cartesian coordinates of the four particles as zero and form the matrix of cartesian coordinates
$$
\left(\begin{array}{cccc}
x_1 & x_2 & x_3 & x_4 \\
y_1 & y_2 & y_3 & y_4 \\
0 & 0 & 0 & 0
\end{array} \right)\, .
$$

The four directed areas are written in terms of these coordinates as
$$
S_1 = \left| \begin{array}{ccc}
1 & 1 & 1 \\
x_2 & x_3 & x_4 \\
y_2 & y_3 & y_4
\end{array} \right| \, \quad S_2 = \left| \begin{array}{ccc}
1 & 1 & 1 \\
x_1 & x_4 & x_3 \\
y_1 & y_4 & y_3
\end{array} \right|
$$
\begin{equation}
S_3 = \left| \begin{array}{ccc}
1 & 1 & 1 \\
x_1 & x_2 & x_4 \\
y_1 & y_2 & y_4
\end{array} \right| \, \quad S_4 = \left| \begin{array}{ccc}
1 & 1 & 1 \\
x_1 & x_3 & x_2 \\
y_1 & y_3 & y_2
\end{array} \right| \, ,
\end{equation}
that are the four signed 3 $\times$ 3 minors formed from the matrix
\begin{equation}
\left(\begin{array}{cccc}
1 & 1 & 1 & 1 \\
x_1 & x_2 & x_3 & x_4 \\
y_1 & y_2 & y_3 & y_4
\end{array} \right)
\end{equation}

Addition to the previous matrix of a row equal to any of its three rows produces a square matrix with zero determinant, implying that the necessary and sufficient conditions to have a flat tetrahedron are
\begin{equation}
\sum_{i=1}^4 S_i = 0\, ,\label{plan}
\end{equation}
and
\begin{equation}
\sum_{i=1}^4 S_i x_i = 0\, , \quad \sum_{i=1}^4 S_i y_i = 0\, .
\end{equation}
The two last equations may be grouped in the zero vector condition
\begin{equation}
\sum_{i=1}^4 S_i {\bf r}_i = {\bf 0}\,  .\label{afin}
\end{equation}

From these properties we deduce a necessary condition for flat solutions in terms of distances.
$$
0 = \sum_{i=1}^4 \sum_{j=1}^4 r_{ij}^2 S_i S_j = \sum_{i=1}^4 \sum_{j=1}^4 ({\bf r}_i^2 - 2 {\bf r}_i \cdot {\bf r}_j + {\bf r}_j^2) S_i S_j =
 $$
 \begin{equation}
 2 (\sum_{i=1}^4 S_i)( \sum_{j=1}^4 {\bf r}_j^2 S_j) - 2 (\sum_{i=1}^4 S_i {\bf r}_i) \cdot (\sum_{j=1}^4  S_j {\bf r}_j)\, .\label{constr}
\end{equation}
This is zero as a consequence of the previous conditions for a flat solution. Indeed the first term is zero because the sum of the directed areas is zero, while the second term is zero by equation (\ref{afin}). Note that equations (\ref{plan} - \ref{constr}) are purely geometrical, independent of the origin of coordinates and independent of the masses. Beside the condition of the sum of directed areas equal to zero, other authors take into account the Cayley-Menger determinant, but we do not. Constraint (\ref{constr}) will be seen to be very important in the theory of central configurations in section 5. Equations (\ref{plan}) and (\ref{afin}) were considered by Dziobec \cite{dz} and are currently connected to affine geometry and the plane four-body problem (see Albouy et al. \cite{al} and references to Albouy therein; note that equation (\ref{constr}) also follows trivially from equations used by Albouy who gives it \cite{alb} in explicit form.)

Other important remark is that although other authors generalize results for $n$ particles in $n-2$ dimensions, and other forms of the potential energy, we do not follow that path.

Implicit in what follows is Heron's formula relating the absolute value of the area of a triangle to the square root of a function of the three sides, which can be expressed, for example, as
\begin{equation}
(4\, S_2)^2 = ( r_{13}^2 \quad r_{43}^2 \quad r_{41}^2 ) \left( \begin{array}{rrr}
-1 & 1 & 1 \\
1 & -1 & 1 \\
1 & 1 & -1
\end{array} \right) \left( \begin{array}{c}
r_{13}^2 \\
r_{43}^2 \\
r_{41}^2
\end{array} \right)\, . \label{heron}
\end{equation}

\section{Central configurations}
In this section we begin with Laplace's approach (see ref. \cite{bp}). Central configurations are defined by the condition that the force divided by the corresponding mass is proportional to the position vector:
\begin{equation}
B {\bf r}_i = \sum_{j \neq i}^4 \frac{m_j ({\bf r}_j -{\bf r}_i)}{r_{ij}^3} \quad \mbox{(for i = 1, 2, 3, 4,)} \label{central}
\end{equation}
where $B$ is a constant, the same for all the particles. Equations (\ref{central}) imply that the origin of coordinates is at the center of mass.

Taking the external product of equations $i$ and $k$ with $({\bf r}_i - {\bf r}_k)\wedge$ and equating, we find the six conditions
\begin{eqnarray}
\label{41} {\bf A}_4 \left( \frac{1}{r_{31}^3} - \frac{1}{r_{12}^3}\right) & = & {\bf A}_1 \left( \frac{1}{r_{43}^3} - \frac{1}{r_{42}^3}\right) \\
{\bf A}_4 \left( \frac{1}{r_{12}^3} - \frac{1}{r_{23}^3}\right) & = & {\bf A}_2 \left( \frac{1}{r_{41}^3} - \frac{1}{r_{43}^3}\right) \\
{\bf A}_4 \left( \frac{1}{r_{23}^3} - \frac{1}{r_{31}^3}\right) & = & {\bf A}_3 \left( \frac{1}{r_{42}^3} - \frac{1}{r_{41}^3}\right) \\
{\bf A}_3 \left( \frac{1}{r_{42}^3} - \frac{1}{r_{12}^3}\right) & = & {\bf A}_2 \left( \frac{1}{r_{43}^3} - \frac{1}{r_{31}^3}\right) \\
{\bf A}_1 \left( \frac{1}{r_{43}^3} - \frac{1}{r_{23}^3}\right) & = & {\bf A}_3 \left( \frac{1}{r_{41}^3} - \frac{1}{r_{12}^3}\right) \\
\label{23} {\bf A}_2 \left( \frac{1}{r_{41}^3} - \frac{1}{r_{31}^3}\right) & = & {\bf A}_1 \left( \frac{1}{r_{42}^3} - \frac{1}{r_{23}^3}\right)
\end{eqnarray}
in terms of the weighted vector areas ${\bf A}_j = {\bf S}_j / m_j$
\begin{eqnarray}
\label{1} {\bf A}_1 & = & \frac{1}{m_1} [{\bf r}_4 \wedge {\bf r}_2 + {\bf r}_2 \wedge {\bf r}_3 + {\bf r}_3 \wedge {\bf r}_4] \\
{\bf A}_2 & = & \frac{1}{m_2} [{\bf r}_1 \wedge {\bf r}_4 + {\bf r}_4 \wedge {\bf r}_3 + {\bf r}_3 \wedge {\bf r}_1] \\
{\bf A}_3 & = & \frac{1}{m_3} [{\bf r}_1 \wedge {\bf r}_2 + {\bf r}_2 \wedge {\bf r}_4 + {\bf r}_4 \wedge {\bf r}_1] \\
\label{4} {\bf A}_4 & = & \frac{1}{m_4} [{\bf r}_2 \wedge {\bf r}_1 + {\bf r}_3 \wedge {\bf r}_2 + {\bf r}_1 \wedge {\bf r}_3]\, .
\end{eqnarray}
These equations are attributed to E. Laura and H. Andoyer (see ref \cite{al}).

In the non-planar case, obviously no pair of these four vectors are parallel for arbitrary masses and positions, and the unique three dimensional central configuration satisfying equations (\ref{41}-\ref{23}) requires all six distances to be the same, so that the masses lie at the vertices of an equilateral tetrahedron. This central configuration has the only solution \cite{bp} of straight paths towards or away from the center of mass. In the rest of the paper we shall consider plane non-collinear central configurations of four masses.

The plane solutions with zero enclosed volume but finite area are obtained by taking into account that the four vectors (\ref{1}-\ref{4}) are now parallel. They are written in terms of the weighted directed areas $A_j$ defined as
\begin{equation}
{\bf A}_j = A_j \bf k\, ,
\end{equation}
with
\begin{equation}
A_j = \frac{S_j}{m_j}\, ,
\end{equation}
where $S_j$ is the value in vector ${\bf S}_j$ multiplying the vector $\bf k$, equal to the directed area of the triangle with the three particles different from $j$ at its vertices.

Suppression of vector $\bf k$ from equations (\ref{41}-\ref{23}) leads us to the homogeneous system
\begin{equation}
\left( \begin{array}{rrrrrr}
0 & A_4 & -A_4 & 0 & A_1 & -A_1 \\
-A_4 & 0 & A_4 & -A_2 & 0 & A_2 \\
A_4 & -A_4 & 0 & A_3 & -A_3 & 0 \\
0 & A_2 & -A_3 & 0 & A_3 & -A_2 \\
-A_1 & 0 & A_3 & -A_3 & 0 & A_1 \\
A_1 & -A_2 & 0 & A_2 & -A_1 & 0
\end{array} \right)
\left( \begin{array}{c}
1/r_{23}^3 \\
1/r_{31}^3 \\
1/r_{12}^3 \\
1/r_{41}^3 \\
1/r_{42}^3 \\
1/r_{43}^3
\end{array} \right) = \left( \begin{array}{c}
0 \\
0 \\
0 \\
0 \\
0 \\
0
\end{array} \right)\, . \label{antis}
\end{equation}
The matrix notation is ours. Since this equation may be considered a consequence of Dziobek's equations, they are frequently attributed to Dziobek. but they were obtained previously, in a similar form to that given here, by E. J. Routh (see \cite{al}.)

Since the coefficient matrix of this system is antisymmetric, a nontrivial solution exists that is the linear combination of its two eigenvectors with eigenvalue zero, namely
$$
( 1 \; 1 \; 1 \; 1 \; 1\; 1) \quad \mbox{and} \quad (A_2 A_3 \; A_3 A_1 \; A_1 A_2 \; A_4 A_1 \; A_4 A_2 \; A_4 A_3) \, .
$$
The first eigenvalue is evident considering the structure of the rows of the matrix and the equilateral solution. The existence of the second eigenvalue is secured because the matrix is antisymmetric of order six. This fact is the consequence that a real antisymmetric matrix becomes hermitian by multiplying its elements by the imaginary unit $i$. The eigenvalues of a hermitian matrix are real. Therefore, the eigenvalues of a real antisymmetric matrix are pure imaginary numbers forming complex conjugate  pairs or zero. If the order of the square matrix is an odd number we have at least one zero eigenvalue. If the order of the antisymmetric matrix is an even number, the zero eigenvalues appear in pairs.

Hence the solution of (\ref{antis}) is given in terms of two parameters $\lambda$ and $\sigma$
\begin{equation}
r_{jk}^{-3} = \sigma + \lambda A_j A_k\, .\label{rjk}
\end{equation}
This equation was also obtained by Dziobek \cite{dz} from a different approach to this problem, considering the critical points of the potential energy, constrained to a constant inertia moment and zero volume (this is the perspective of ref. \cite{ds}). One way to take into account the zero volume condition is by taking the derivative with respect to $r_{ij}^2$ of the Cayley-Menger determinant (which is proportional to the square of the volume of the tetrahedron). These derivatives are proportional to the product $S_i S_j$ if the plane restrictions are taken in account. An obvious difficulty of that deduction is that since the non trivial entries of the Cayley Menger determinant are just the squares $r_{ij}^2$, to obtain the square roots in the expression of the directed areas $S_i$ as functions of the $r_{ij}^2$, equations (\ref{heron}), one needs to use explicitly restrictions (\ref{plan}) and (\ref{afin}) connecting these areas. Actually Dziobek's proof remained unpublished until the end of the Twentieth Century (see Moeckel \cite {mo}). Other proof using convex theory has been developed by Albouy (see \cite{al} and references therein). Although D. Saari follows Dziobek's approach of critical points, he does not recommend \cite{sa} the use of Cayley-Menger determinant but instead he is in favor of geometric arguments. Note that our proof is quite elementary using simple properties of linear algebra and we follow Saari's recommendation of geometric arguments \cite{sa}, using Dziobek's geometric restrictions of the previous section (\ref{plan}) and (\ref{afin}), also made explicit in Albouy et al. \cite{al}.

Now we obtain a well known expression for the parameter $\sigma$, which starts by writing (\ref{rjk}) in the form
\begin{equation}
m_j m_k r_{jk}^{-3} = m_j m_k \sigma + \lambda S_j S_k\, .\label{dziobek}
\end{equation}
Multiply both sides by $r_{jk}^2$ and sum over $j$ and $k$. Using the geometric plane constraint (\ref{constr}) we find
\begin{equation}
\sigma = \frac{\sum\limits_{j > k} m_j m_k /r_{jk}}{\sum\limits_{j > k} m_j m_k r_{jk}^2}\, , \label{sigma}
\end{equation}
expression that is positive definite. From either (\ref{sigma}) or (\ref{rjk}) it is seen that $\sigma$ has dimensions of the reciprocal of a cubed length.

Further, substitution of (\ref{rjk}) in the equation that defines central configurations (\ref{central}), we verify that it holds with the constant $B$ given by
\begin{equation}
B = - \sigma \sum_{i=1}^4 m_i\, .
\end{equation}
This just requires that the origin of coordinates is located at the center of mass and the plane constraints (\ref{plan}) and (\ref{afin}).

\begin{figure}
\centering
\scalebox{0.5}{\includegraphics{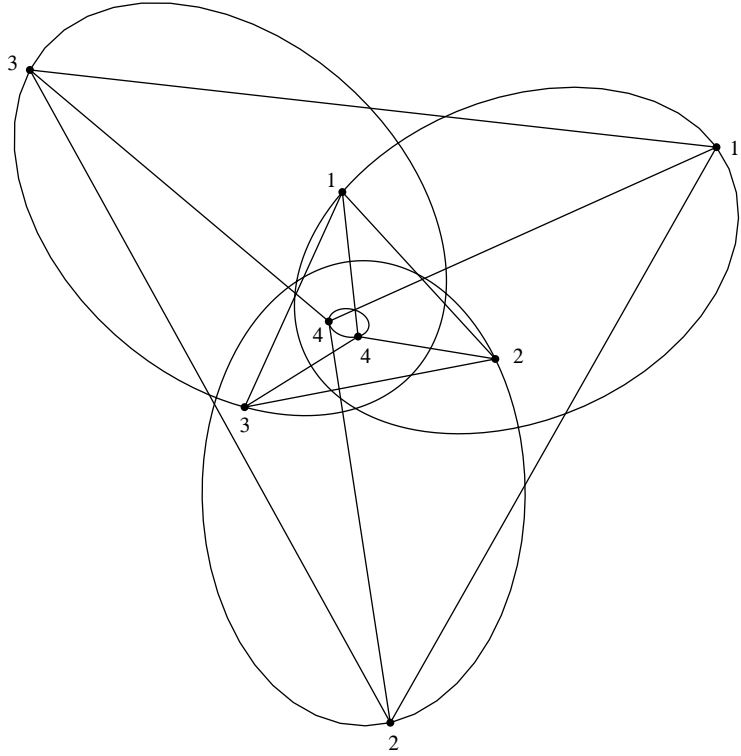}}

\caption{Two different arrangements of four particles with different masses following elliptic trajectories in a central concave configuration. The common focus of the four ellipses is at the right hand side focus of the small ellipse. The eccentricity of the four ellipses is $e=0.72$. $A_1 = 5$, $A_2 = 6$, $A_3 = 4$, $A_4 = -8$.}
\end{figure}

On the other hand, we have that $\lambda$ is negative. For the proof, we write explicitly the six equations in (\ref{rjk}) as
\begin{eqnarray}
\label{33} r_{12}^{-3} - \sigma & = & \lambda A_1 A_2 \\
r_{23}^{-3} - \sigma & = & \lambda A_2 A_3 \\
\label{35} r_{31}^{-3} - \sigma & = & \lambda A_3 A_1 \\
\label{36} r_{41}^{-3} - \sigma & = & \lambda A_4 A_1 \\
\label{37} r_{42}^{-3} - \sigma & = & \lambda A_4 A_2 \\
\label{38} r_{43}^{-3} - \sigma & = & \lambda A_4 A_3 \, ,
\end{eqnarray}
and for convenience we consider separately the concave and convex cases. The concave case occurs when one particle is in the convex hull of the other three. In the convex case no particle is in the convex hull of the other three. Instead of formal definitions we appeal to two general examples of concave and convex configurations represented by our drawings in Fig. 1 and Fig. 2.

\begin{figure}
\centering
\scalebox{0.5}{\includegraphics{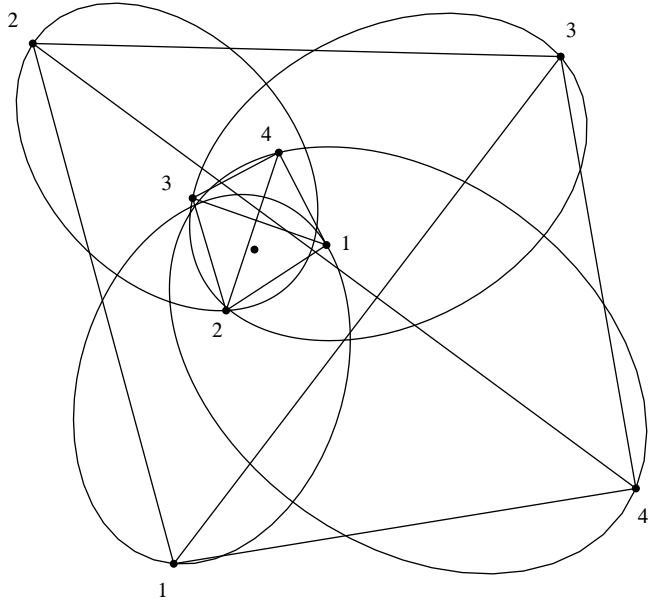}}

\caption{Two different arrangements of four particles with different masses following elliptic trajectories in a central convex configuration. The isolated point is the center of mass and the common focus of the four ellipses. The eccentricity of the four ellipses is $e=0.72$. $A_1 = 15$, $A_2 = -6$, $A_3 = 3$, $A_4 = -4$.}
\end{figure}

\noindent A) First the concave case where a particle, labeled 4, is in the convex hull of the other three, labeled 1, 2, 3.
Assume with no loss of generality that
\begin{equation}
A_1 \geq A_2 \geq A_3 > 0 > A_4\, . \label{desig}
\end{equation}
Then the first three equations (\ref{33}-\ref{35}) have the sign of $\lambda$ and the last three equations (\ref{36}-\ref{38}) have the opposite sign. Assume $r_{ij}$ is the largest of $(r_{12}, r_{23}, r_{31})$, then the equilateral triangle based on $r_{ij}$ overlaps triangle 123. Therefore, since mass 4 is in the convex hull of the others, distances $r_{4i} < r_{ij}$ and hence $r_{4i}^{-3} - \sigma > r_{ij}^{-3} - \sigma$. Since one of these is positive and the other negative, we can only have
\begin{equation}
r_{4i}^{-3} - \sigma > 0 \, , \quad r_{ij}^{-3} - \sigma < 0 \, ,
\end{equation}
from which follows that
\begin{equation}
\lambda < 0 \, .
\end{equation}
As a Corollary of the previous proof, distances $r_{12}, r_{23}, r_{31}$ are larger than $\sigma^{-1/3}$, which is larger than $r_{41}, r_{42}, r_{43}$.

\noindent B) Second, the convex case where no particle is in the convex hull of the other three. We label the ends of the two diagonals with numbers 1, 3 and 2, 4, respectively. The four labels are in the cyclic order 1234. Assume with no loss of generality
\begin{equation}
A_1 \geq A_3 > 0 > A_2 \geq A_4 \, .
\end{equation}
Thus equations (\ref{35}) and (\ref{37}) have the sign of $\lambda$ and the other four of the set (\ref{33}-\ref{38}) have the opposite sign. Assume $r_{ij}$ is the longest diagonal ($r_{13}$ or $r_{24}$) and $r_{ik}$ is the shortest side of the set ($r_{12}, r_{23}, r_{43}, r_{41}$.) Then $r_{ij} > r_{ik}$ since assuming the contrary the four sides would be longer than $r_{ij}$, and then the two equilateral triangles with common side $r_{ij}$ could be covered by the quadrangle 1234 and then $r_{ij}$ would not be the longest diagonal, contrary to the hypothesis. Then $r_{ij}^{-3} - \sigma < r_{ik}^{-3} - \sigma$ , but since they must be of opposite signs we must have
\begin{equation}
r_{ij}^{-3} - \sigma < 0 \, , \quad r_{ik}^{-3} - \sigma > 0 \, ,
\end{equation}
from which
\begin{equation}
\lambda < 0\, .
\end{equation}
As a corollary of this proof, the diagonals are larger than $\sigma^{-1/3}$ which is larger than the longest of the four sides.

  A different proof that $\lambda$ is negative is obtained in \cite{al} as corollary of lemma 1. The sign of $\lambda$ is important because numerical solutions to equation (\ref{antis}) of the form (\ref{rjk}) may be found with $\lambda$ positive, that obey restriction (\ref{plan}) but are not planar central configurations since they do not satisfy (\ref{afin}) nor (\ref{constr}).

Additional relations between distances are obtained from the quotient, member by member, of equations (\ref{33}-\ref{38}). One has
\begin{eqnarray}
\label{43} \frac{A_1}{A_3} & = & \frac{r_{12}^{-3} - \sigma}{r_{23}^{-3} - \sigma} = \frac{r_{41}^{-3} - \sigma}{r_{43}^{-3} - \sigma} = \frac{r_{12}^{-3} - r_{41}^{-3}}{r_{23}^{-3} - r_{43}^{-3}} \\
\label{44} \frac{A_2}{A_3} & = & \frac{r_{12}^{-3} - \sigma}{r_{31}^{-3} - \sigma} = \frac{r_{42}^{-3} - \sigma}{r_{43}^{-3} - \sigma} = \frac{r_{12}^{-3} - r_{42}^{-3}}{r_{31}^{-3} - r_{43}^{-3}} \\
\frac{A_4}{A_3} & = & \frac{r_{42}^{-3} - \sigma}{r_{23}^{-3} - \sigma} = \frac{r_{41}^{-3} - \sigma}{r_{31}^{-3} - \sigma} = \frac{r_{42}^{-3} - r_{41}^{-3}}{r_{23}^{-3} - r_{31}^{-3}} \\
\label{46}\frac{A_1}{A_2} & = & \frac{r_{31}^{-3} - \sigma}{r_{23}^{-3} - \sigma} = \frac{r_{41}^{-3} - \sigma}{r_{42}^{-3} - \sigma} = \frac{r_{31}^{-3} - r_{41}^{-3}}{r_{23}^{-3} - r_{42}^{-3}} \\
\frac{A_4}{A_2} & = & \frac{r_{41}^{-3} - \sigma}{r_{12}^{-3} - \sigma} = \frac{r_{43}^{-3} - \sigma}{r_{23}^{-3} - \sigma} = \frac{r_{41}^{-3} - r_{43}^{-3}}{r_{12}^{-3} - r_{23}^{-3}} \\
\frac{A_4}{A_1} & = & \frac{r_{42}^{-3} - \sigma}{r_{12}^{-3} - \sigma} = \frac{r_{43}^{-3} - \sigma}{r_{31}^{-3} - \sigma} = \frac{r_{42}^{-3} - r_{43}^{-3}}{r_{12}^{-3} - r_{31}^{-3}} \, ,
\label{48} \end{eqnarray}
which are independent of the parameter $\lambda$. The last expression on the right hand side of these equations results from the previous two, assuming that they are different. The middle terms imply restrictions that may be also obtained from the original equations, namely
\begin{equation}
(r_{12}^{-3} - \sigma)(r_{43}^{-3} - \sigma) = (r_{31}^{-3} - \sigma)(r_{42}^{-3} - \sigma) = (r_{23}^{-3} - \sigma)(r_{41}^{-3} - \sigma) \, ,
\end{equation}
which all equal $ \lambda^2 A_1 A_2 A_3 A_4$.

From these equations one also obtains the $\sigma$ parameter in terms of the distances
$$
\sigma = \frac{r_{12}^{-3} r_{43}^{-3} - r_{23}^{-3} r_{41}^{-3}}{r_{12}^{-3} + r_{43}^{-3} - r_{23}^{-3} - r_{41}^{-3}} = \frac{r_{31}^{-3} r_{42}^{-3} - r_{12}^{-3} r_{43}^{-3}}{r_{31}^{-3} + r_{42}^{-3} - r_{12}^{-3} - r_{43}^{-3}} =
$$
\begin{equation}
 \frac{r_{23}^{-3} r_{41}^{-3} - r_{31}^{-3} r_{42}^{-3}}{r_{23}^{-3} + r_{41}^{-3} - r_{31}^{-3} - r_{42}^{-3}}\, .\label{54}
\end{equation}
Indeed this result is an identity after substituting the reciprocal cubed distances (\ref{rjk}), in the right hand side of this equation.

All of equations (\ref{43}-\ref{54}) appear in Schmidt \cite{ds}. We reproduce them here because they will be useful to prove further important results.

Coming back to the concave case we assume now the inequalities (cf. equation (\ref{desig})
\begin{equation}
A_1 > A_2 > A_3 > 0 > A_4 \, .
\end{equation}
Then equations (\ref{43}-\ref{48}) imply
\begin{equation}
r_{12} > r_{31} > r_{23} > \sigma^{-1/3} > r_{43} >  r_{42} >  r_{41}\, .
\end{equation}

Returning to the convex case, now we assume
\begin{equation}
A_1 > A_3 > 0 > A_2 > A_4\, .
\end{equation}
Equations (\ref{43}-\ref{48}) then yield two inequalities for the four sides
\begin{equation}
\sigma^{-1/3} > r_{23} > r_{12} > r_{41}\, , \quad \sigma^{-1/3} > r_{23} > r_{43} > r_{41}\, ,
\end{equation}

\section{Symmetric configurations}
Important symmetric cases are also deduced from equations (\ref{43}-\ref{48}).

The necessary and sufficient condition for a kite symmetry is the equality of two weighted areas $A$'s, for example iff
\begin{equation}
A_1 = A_3 \,\label{57}
\end{equation}
then
\begin{equation}
r_{12} = r_{23} \quad \mbox{ and } \quad r_{41} = r_{43}
\end{equation}
and, furthermore,
\begin{equation}
S_1 = S_3 \quad \mbox{ and }\quad m_1 = m_3\, .\label{59}
\end{equation}
The kite symmetry imposes the above restrictions. The values of the other weighted directed areas $A_2$ and $A_4$ are arbitrary, but at least one of them must be of opposite sign to that of $A_1$.

Imposing the previous symmetry twice leads us to two important symmetric cases.

\noindent 1)Equilateral triangle symmetry occurs iff
\begin{equation}
A_1 = A_2 = A_3\, ,
\end{equation}
which implies
\begin{equation}
r_{23} = r_{31} = r_{12} > r_{41} = r_{42} = r_{43}\, .
\end{equation}
The three particles 123 are at the vertices of an equilateral triangle and particle 4 is at the center of this triangle, so that
\begin{equation}
S_1 = S_2 = S_3 = - S_4/3\, , \quad m_1 = m_2 = m_3\, , \quad \frac{r_{12}}{r_{41}} = \sqrt{3} \, ;
\end{equation}
with $A_4$ arbitrary, but satisfying the important equation
\begin{equation}
\frac{A_4}{A_1} = - 3 \frac{m_1}{m_4}\, .
\end{equation}

\noindent 2) Rhomboidal symmetry occurs iff
\begin{equation}
A_1 = A_3 > 0 > A_2 = A_4\, ,
\end{equation}
implying that
\begin{equation}
r_{12} = r_{23} = r_{43} = r_{41}\, , \, \mbox{ and }\, S_1 = S_3 = - S_2 =  -S_4\, .
\end{equation}
In order to yield the square symmetry, the diagonals must be equal, requiring the extra condition
\begin{equation}
A_1 A_3 = A_2 A_4\, .
\end{equation}

\noindent 3) Another kind of symmetry is obtained iff
\begin{equation}
A_1 = - A_2 \quad \mbox{ and }\quad  A_3 = - A_4
\end{equation}
which imply the isosceles trapezium symmetry:
\begin{equation}
r_{31} = r_{42} \quad \mbox{ and } \quad r_{23} = r_{41} \label{trapezium}
\end{equation}
since
\begin{equation}
A_1 A_3 = A_2 A_4 \quad \mbox{ and }\quad  A_2 A_3 = A_1 A_4\, .\label{double}
\end{equation}
In such a case of trapezium isosceles symmetry we also find that
\begin{equation}
S_1 = - S_2 \quad \mbox{ and }\quad  S_3 = - S_4\, ,
\end{equation}
which imply
\begin{equation}
m_1 = m_2 \quad \mbox{ and } \quad m_3 = m_4\,.
\end{equation}
Note that assuming the trapezium isosceles symmetry through (\ref{trapezium}), equation (\ref{double}) follows with the possible symmetries of the rhombus or of the isosceles trapezium.

The isosceles trapezium symmetry gives the stronger square symmetry if the four $A_j$'s have the same absolute value.

\section{A new algorithm}
Equations (\ref{plan}), (\ref{heron}) and (\ref{dziobek}) have been often used by several authors to compute the six distances $r_{jk}$ from given values of the four masses. According to D. Saari \cite{sa} this perspective makes the problem difficult to manipulate. For example, with this perspective in mind, some advances for the case of three equal masses with kite symmetry have been published by Bernat et al. \cite{be}, with important results about bifurcation values corresponding to different number of solutions. With our approach the problem with two equal masses is simpler than the Bernat et al. paper where the restriction of three equal masses leads to a more complex algorithm.

Since the lengths and masses are defined up to arbitrary units and noting that the parameter $\sigma$ has units of the inverse of a volume, we assume with no loss of generality that all the distances are measured in the unit of distance $\sigma^{-1/3}$ which simplifies equation (\ref{rjk}) into
\begin{equation}
r_{jk}^{-3} = 1 + \lambda A_j A_k\, .
\end{equation}
In the following, we shall therefore use the simplification $\sigma = 1$, and we also assume that the directed weighted areas $A_k$ are known as four given constants (the arbitrariness of the four values is justified by the arbitrariness of the four masses.) The previous equation gives all the distances as functions of the unknown parameter $\lambda$ and the four constants $A_k$. From them, using Heron's formula (\ref{heron}), the areas of the four triangles become functions of $\lambda$. The sign of a directed area $S_j$ is inherited from the same sign of the corresponding constant $A_j$. Restriction (\ref{constr}) then determines the value of $\lambda$ for a plane solution and from it, the value of the six distances and the four masses are found (in terms of the four constants $A_j$). This is an implicit way to deduce planar central configurations with four masses. In many non symmetric cases restriction (\ref{plan}) is sufficient to define a central configuration, but in a few cases, restriction (\ref{constr}) is not automatically satisfied and allows one to discriminate non-physical situations (such as negative distances or evidently non-planar solutions or geometrically impossible figures.) We stress the fact that to determine $\lambda$, we chose to find the root of (\ref{constr}) as a function of $\lambda$. In any case considered numerically this solution is unique.

This approach to the four body central configurations is similar (see \cite{sa}) to considering the Euler collinear central configurations of three particles, where instead of giving the ratio of the masses and computing the corresponding ratio of two distances with a quintic equation, the ratio of two distances is first given and the ratio of the masses is computed from just a linear equation.

The symmetric cases of kite symmetry, equilateral triangle symmetry, rhomboidal symmetry, and trapezium isosceles symmetry were presented from this approach, based on the weighted directed areas $A_j$.

In general, given a solution, it is invariant to a change of sign of the four constants $A_j$. Therefore, with no loss of generality we may assume $A_1, A_3$ positive and $A_4$ negative, henceforth the concave case is determined by $A_2$ positive and the convex case by $A_2$ negative.

More generally, consider that a solution with a set of values of the four constants $A_j$ is determined by the value of $\lambda$. Multiplying the four constants by a real $k$: $A_j \longrightarrow k A_j$ the solution requires a transformed $\lambda$ as: $\lambda \longrightarrow \lambda/k^2$.

This procedure produces in most cases a solution, but we find some cases where values of these constants yield none. Examples are given in the next section devoted to increase our study of the kite symmetry.

\section{Asymptotic cases of Kite Symmetry}
In order to deduce more of its properties, we come back to the kite symmetry recalling equations (\ref{57}-\ref{59}) that we rewrite here
$$
A_1 = A_3 \, ,
$$
then
$$
r_{12} = r_{23} \quad \mbox{ and } \quad r_{41} = r_{43}
$$
and further
$$
S_1 = S_3 \quad \mbox{ and }\quad m_1 = m_3\, .
$$

We shall consider four asymptotic cases of kite symmetry where one of the four masses approaches either zero or infinity. That mass lies on the symmetry axis between masses $m_1$ and $m_3$. For clarity we subdivide these four cases in different subsections.

We discuss first the concave solution with $A_1 = A_3$, $A_2$ positive and $A_4$ negative. $m_4$ is in the convex hull of the other three particles. Particles with masses $m_2$ and $m_4$ are placed on the symmetry axis of particles with equal masses $m_1$ and $m_3$.

Instead of the planar condition (\ref{constr}) we use Pythagoras' Theorem which yields the equivalent equation
\begin{equation}
r_{42}-\sqrt{r_{12}^2 -\frac{r_{31}^2}{4}} +\sqrt{r_{41}^2 -\frac{r_{31}^2}{4}} = 0 \, .\label{pitag}
\end{equation}
This constraint on the distances between particles is simpler than the other condition, resulting in a smaller numerical error, although it is only valid for the concave kite symmetry.

\subsection{Euler convex}
Numerical evidence was obtained considering fixed values of $A_1$ and $A_4$, and $A_2$ variable. Starting with $A_2 = A_1 = A_3$, the equilateral triangle solution, we decrease only the $A_2$ value. We observed that as $A_2$ is decreased, $m_2$ approaches zero until an asymptotic value of $A_2$ is reached where there is no solution. Since $A_2$ is a positive number, and the mass $m_2$ approaches zero, this implies that the directed area $S_2$ also tends to zero, but this is possible only when $r_{41}$ approaches the value $r_{31}/2$, which was verified numerically. Particles 1, 4, 3 become asymptotically collinear, with particle 4 at the midpoint of $r_{13}$. Imposing the equality $r_{41} = r_{31}/2$ the allowed limiting value of $\lambda$ is found to be
\begin{equation}
\lambda = - \frac{7}{8 A_1^2 - A_1 A_4}\, ,
\end{equation}
a function of $A_1$ and $A_4$ only. Substitution in the constraint (\ref{pitag}) as function of the constants $A_j$ and $\lambda$ (that becomes the Pythagoras Theorem of the right triangle formed by 1, 2, 4,) one deduces the limiting minimum value of $A_2/A_1$ as the root $x$ of the equation
\begin{equation}
\frac{1}{(8 A_1 - A_4 - 7 A_1 x)^\frac{2}{3}} - \frac{1}{(8 A_1 - A_4 - 7 A_4 x)^\frac{2}{3}} = \frac{1}{(8 A_1 - 8 A_4)^\frac{2}{3}} \, .
\end{equation}
This root is a function of the ratio $-A_4/A_1$ only. As this ratio approaches zero, the limit of $A_2/A_1$ tends to
\begin{equation}
\frac{1}{7} \left(8 - \sqrt{8} \right) = 0.738796125... \, .
\end{equation}
On the other hand, as the ratio $-A_4/A_1$ approaches $\infty$, the limit of $A_2/A_1$ tends to
\begin{equation}
\frac{1}{7} \left(\frac{8}{\sqrt{27}} - 1 \right) = 0.077085817...\, .
\end{equation}
In this limit, as the mass $m_2$ tends to zero, the other three particles tend to lie on a straight line, a central configuration first studied by Euler.

\subsection{Lagrange concave}
In the second asymptotic case, we again start from the equilateral solution with $A_1 = A_3 = A_2$, all positive, and a fixed $A_4<0$, but now we increase the value of $A_2$. As $A_2$ grows to infinity the three distances $r_{41}, r_{43}, r_{31}$, approach 1, $m_2$ tends to zero, $r_{12} = r_{23}$ tend to a limit larger than 1, $r_{42}$ has a limit smaller than 1, and $\lambda$ approaches zero from below. This limit for $\lambda$ implies that $r_{41}, r_{43}, r_{31}$ tend to 1. The peculiar behavior of $r_{12} = r_{23}$ and $r_{42}$ is explained by the property that
\begin{equation}
\lim_{\lambda\to -0, A_2\to\infty} \lambda A_2 = \mbox{finite number } < 0\, .
\end{equation}

To compute the unknown distances $r_{42}$ and $r_{12}$ we may apply the first right hand side of equation (\ref{48}) with $\sigma = 1$:
$$
\frac{A_4}{A_1} = \frac{r_{42}^{-3} - 1}{r_{12}^{-3} - 1}\, .
$$
In this limiting condition, the angle 143 is $\pi/3$ and angle 241 is $5 \pi/6$. The trigonometric cosine-sides relation for triangle 241 gives in the limit case the additional relation
\begin{equation}
1 + r_{42}^2 + r_{42} \sqrt{3} = r_{12}^2 \, .
\end{equation}
Eliminating $r_{12}$ from these two equations we obtain an equation for $r_{42}$ that is a function of the ratio $A_4/A_1$
\begin{equation}
\frac{A_4}{A_1} = \frac{r_{42}^{-3} - 1}{(1+ r_{42}^2 + r_{42} \sqrt{3})^{-3/2} - 1}\, .
\end{equation}
In this case as $A_2 \rightarrow \infty$, the mass $m_2$ tends to zero. The other three particles are in an equilateral triangle central configuration first studied by Lagrange. From this value of $r_{42}$, the limiting value $\lambda A_2$ may be computed from equation (\ref{37}) as
$$
\lambda A_2 = \frac{r_{42}^{-3} - 1}{A_4}\, .
$$

\subsection{Lagrange convex}
We now turn to the convex solution with $A_1$ = $A_3$ positive and $A_2$, $A_4$ negative, no particle is in the convex hull of the other three. Just as in the concave case, particles with masses $m_2$ and $m_4$ are on the symmetry axis of the particles with equal masses $m_1$ and $m_3$.

The Theorem of Pythagoras in this case gives us the constraint
\begin{equation}
r_{42}-\sqrt{r_{12}^2 -\frac{r_{31}^2}{4}} - \sqrt{r_{41}^2 -\frac{r_{31}^2}{4}} = 0 \, ,
\end{equation}
valid only for the convex configuration with kite symmetry.

We have studied numerically this convex case of kite symmetry keeping the two positive weighted areas $A_1$ and $A_3$ and the negative weighted area $A_2$ constant and tabulating the six distances and four masses when different values of the
weighted negative area $A_4$ are given. Any chosen negative value of this constant gives a solution.

In the limit as $A_4$ tends to minus infinity, $\lambda$ goes to  minus zero with the property
\begin{equation}
\lim_{\lambda\to -0, A_4\to-\infty} \lambda A_4 = \mbox{finite number } > 0\, .
\end{equation}
In this limit, the mass 4 approaches zero and the other three particles form a Lagrange's equilateral solution with the three distances $r_{12}$, $r_{23}$, $r_{31}$ tending to the value 1. The unknown distances $r_{41}$ and $r_{42}$ may be calculated in similar form to the convex Lagrange case. The two distances are related by equation (\ref{46})
$$
\frac{A_1}{A_2} = \frac{r_{41}^{-3} - 1}{r_{42}^{-3} - 1}\, .
$$
Again a trigonometric relation between the two distances is obtained from the triangle 124. Angle 124 is $\pi/6$; the cosine-sides relation for this triangle produces
\begin{equation}
1 + r_{42}^2 - r_{42} \sqrt{3} = r_{41}^2\, .
\end{equation}
Elimination of $r_{41}$ between the last two equations gives side $r_{42}$ as a function of the ratio $A_1 / A_2$, as the root of the equation
\begin{equation}
\frac{A_1}{A_2} = \frac{(1 + r_{42}^2 - r_{42} \sqrt{3} )^{-3/2} - 1}{r_{42}^{-3} - 1}\, .
\end{equation}

With this value of $r_{42}$, the finite limiting value of $\lambda A_4$ may be computed now from equation (\ref{37})
$$
\lambda A_4 = \frac{r_{42}^{-3} - 1}{A_2}\, .
$$

\subsection{Coorbital}
Turning presently to the examination of the convex solution with $A_1$ = $A_3$ positive and $A_2$, $A_4$ negative in the limit where the negative constant $A_4$ tends to zero, we find that the three distances $r_{41}, r_{42}, r_{43}$ approach the same limit 1. In this limit, the three particles 1, 2, 3 lie on a circle with center at the position of particle 4, with a mass tending to infinity. This corresponds physically to the coorbital case of particles of negligible mass, around a central large mass, whose study was pioneered by Maxwell in 1856. See for example reference \cite{ll} for a panoramic view of this problem.

The kite symmetry implies $\theta =$ angle 142 and that $ 2 \theta = $angle 143. Therefore the two distances $r_{12}$ and $r_{13}$ are given by
\begin{equation}
\sin \frac{\theta}{2} = \frac{r_{12}}{2} \, , \quad \sin \theta = \frac{r_{31}}{2} \, .
\end{equation}
They are related by the trigonometric identity
\begin{equation}
r_{31} = 2 r_{12} \sqrt{1 -r_{12}^2/4}
\end{equation}

Distances $r_{12}$ and $r_{31}$ satisfy also the equation (\ref{44}) with $\sigma = 1$
$$
\frac{A_2}{A_3} = \frac{r_{12}^{-3} - 1}{r_{31}^{-3} - 1}\, .
$$

Eliminating $r_{31}$ between the two last equations allows us to determine $r_{12}$ from the equation
\begin{equation}
\frac{A_2}{A_3} = \frac{r_{12}^{-3} - 1}{(2 r_{12} \sqrt{1 -r_{12}^2/4})^{-3} - 1}\, .
\end{equation}

The limiting value of $\lambda$, as $|A_4| \rightarrow 0$, is deduced from (\ref{33})
$$
\lambda = \frac{r_{12}^{-3} - 1}{A_1 A_2} \, .
$$

In comparison, the approach used in \cite{ll} is to assume the restriction that the three satellite masses are equal. Then two non trivial solutions are given by those authors in terms of angle $\theta$, one corresponding to the convex case, and the other to the concave case. Actually, we find that these angles should both obey the trigonometric restriction
\begin{equation}
1-\frac{1}{8 |\sin^3(\theta/2)|} + 2 \cos(\theta) \left[1 - \frac{1}{8 |\sin^3(\theta)|} \right] = 0 \, ,
\end{equation}
with the two non trivial solutions \cite{ll}
$$
\theta = 0.826602936080376..\, , \quad \theta = 2.4219145305912.. \, .
$$
To relate with our approach we stress that we start with the values for the weighted directed areas
$$
A_1 = A_3 = 1\, , \quad A_4 \dot{=} 0\, , \quad A_2 = \frac{r_{12}^{-3} - 1}{r_{31}^{-3} - 1} = \frac{(2 \sin(\theta/2)^{-3} - 1}{(2 \sin \theta)^{-3} - 1} \, .
$$
Then, computing the masses we found that masses $m_1$=$m_3 \dot{=} m_2$, as might have been expected. Thus the relation of our method with \cite{ll} is established for the (1 + 3) central configuration.

\section{Four different masses}
In the previous section, several limiting cases were presented where some symmetric limit central configurations were computed from simple expressions and compared with numerical experiments.

In the general case of four different masses we have not much theory to add to the previous existing Dziobek's \cite{dz} findings. We have nevertheless to offer the practical algorithm presented in this paper where the four constants $A_j$ are assumed known and from them the distances, and hence the areas, are known as functions of $\lambda$ that are determined from a plane constraint. At the next stage, the four masses are determined instead of giving them from the beginning.

The case with four different masses and no evident symmetry has also asymptotic limiting configurations similar to those cases described in the previous section supported by symmetry and is dealt with using the  same new algorithm.

First the case where one of the weighted directed areas, say $A_k$ tends to infinity, and simultaneously the $\lambda$ parameter goes to zero. The three distances that are expressed in terms of three finite, non-zero, values of the directed weighted areas approach unity, because of the $\lambda$ behavior. They may be assumed 1 from the beginning.  Introducing a new finite parameter equal to the formal product $\lambda A_k$, this is now determined from the plane constraint. This limit corresponds to the physical situation in which one mass disappears and the other three become a Lagrange's equilateral three body central configuration, which is completely determined by the equilateral condition.

The other central configuration of three bodies, the Euler collinear configuration, occurs again in the limit where the fourth mass tends to zero. In this case both the mass and the area associated to the vanishing particle approach zero; the quotient, the corresponding weighted directed area takes a finite non-zero value. To compute the limiting case we assume that only three weighted directed areas of the three collinear particles are known, then introduce the collinear condition on the three distances, function of $\lambda$, between the collinear particles in order to obtain the $\lambda$ value. That this is the usual Euler condition comes from Albouy's proof \cite{al} of Dziobek's equation (\ref{rjk}) that is also valid for the collinear three body case with the directed weighted areas replaced by directed weighted distances. Because the three triangles have the same height, directed areas and directed distances are proportional. The unknown value of the fourth weighted directed area is determined from the planar constraint (\ref{constr}) because it is the only remaining unknown in the expressions for the six distances. This limiting central configuration can be obtained from a concave solution when one of the weighted directed areas changes its value. The deduced limiting value of $A_4$ bounds this weighted directed area from further change.

The coorbital case when one of the masses tends to infinity is obtained in the zero limit of one particular weighted directed area, assuming the other three weighted directed distances are known, finite and non zero. Then the three distances connecting the massive particle become equal to one. We may replace this value from the beginning. The other three weighted directed areas are known in terms of $\lambda$, which is determined as usual. These three distances determine the geometric angles separating the three satellites (with three different masses.)

In the generic case of different masses, the four assumed known weighted directed areas have been related to one of the coordinates of the four particles determined by the internal rotation around the center of mass of a rigid tetrahedron, which is a function of the four masses only, having equal principal moments of inertia. This is the kernel of the coordinates introduced by Pi\~na to describe the general dynamics of the Four-Body Problem \cite{p4} that will be presented in a separate paper. As this kernel rotates (by three Euler angles) it then collapses into a plane, loosing the four coordinates in the direction orthogonal to the plane. These four coordinates, function of the masses and two Euler angles, form four quantities that are proportional to the four weighted directed areas. We refer to that paper for details about these coordinates.

\end{document}